\documentclass[aps,twocolumn,prb]{revtex4-1}
\usepackage{amsfonts}
\usepackage[final,hiresbb]{graphicx}
\usepackage{amsmath}
\usepackage{amssymb}
\usepackage{amstext}
\usepackage{braket}
\usepackage[colorlinks,allcolors=blue]{hyperref}

\usepackage[mathscr]{euscript}

\usepackage{color}

\definecolor{Green}{RGB}{0,204,102}
\definecolor{Purple}{RGB}{102,0,255}
\definecolor{Blue}{RGB}{51,153,255}
\definecolor{Red}{RGB}{151,010,010}

\begin{document}

\title{The Centroid Shifts of Light Beams Reflected from Multi-Layers\\and the Effects of Angular Momentum Manifestations}  
\date{September 21, 2016}

\author{Mark T. Lusk}
\email{mlusk@mines.edu}
\affiliation{Department of Physics, Colorado School of Mines, Golden, CO 80401, USA}
\author{Mark E. Siemens}
\affiliation{Department of Physics \& Astronomy, University of Denver, Denver, CO 80208-6900, USA}
\author{G. F. Quinteiro}
\affiliation{Departamento de F\'isica and IFIBA, FCEN, Universidad de Buenos
Aires, Ciudad Universitaria, Pabell\'on I, 1428 Ciudad de Buenos Aires,
Argentina}

\begin{abstract}
Laguerre-Gaussian (LG) beams reflected from a multi-layered dielectric experience a shift in their centroid that is different than that from a single interface. This has been previously investigated for linearly polarized beams and, to a much lesser extent, beams with spin angular momentum. Here a combination of perturbation and computational analyses is used to provide a unified quantification of these shifts in layered dielectrics with two parallel interfaces. The approach is then extended to consider the qualitatively new behavior that results when the light is endowed with an intrinsic orbital angular momentum--i.e. vortex beams. Destructive interference causes singular lateral shifts in the centroid of the reflected vortex beam for which spin alone produces only a mild modulation. In the case of total internal reflection, both spin and intrinsic orbital angular momentum contribute to an enhancement of these lateral shifts as the interlayer thickness is decreased. This is just the opposite of the trend associated with longitudinal shifts. Two geometries are considered: air/glass/air and glass/air/glass multi-layers. A commonly available glass is used to show that vortex beams can result in centroid shifts on the order of microns for beams with a significant reflection coefficient.  

\end{abstract}
\maketitle

%
\section{Introduction}

It is well known that the centroid of a beam of light, unlike plane waves, can exhibit both longitudinal and lateral shifts when reflected from a dielectric interface. Centroid displacement perpendicular to the beam axis but in the plane of incidence (longitudinal direction) has come to be known as a Goos-H{\"a}nchen (GH) shift~\cite{Goos_Hanchen_1947}. This occurs in association with the total internal reflection of light and is due to the dispersive nature of the complex-valued reflection coefficient, often described in terms of the interfacial propagation of evanescent modes. Displacements perpendicular to the beam axis but out of the plane of incidence (lateral direction) are collectively known as Imbert-Fedorov (IF) shifts~\cite{Fedorov_1955,Fedorov_Translation_2012, Imbert_1972}. The effect is actually attributable to two processes that conserve angular momentum while creating an extrinsic orbital angular momentum (OAM) from spin and/or intrinsic orbital manifestations~\cite{Oneil_2002}. The transfer of spin angular momentum (SAM) to extrinsic OAM can be explained as a photonic version of the Spin-Hall Effect and so is referred to as a Spin-Hall Effect of Light (SHEL) shift~\cite{Onoda_2004}. Momentum transfers from intrinsic to extrinsic OAM, implying that the beam is no longer centrosymmetric about the axis associated with simple reflection, will be referred to as Orbital Imbert-Fedorov (OIF) shifts~\cite{Fedoseyev_1988, Fedoseyev_2001}.  

%
\begin{figure}[hptb]
\begin{center}
\includegraphics[width=0.48\textwidth]{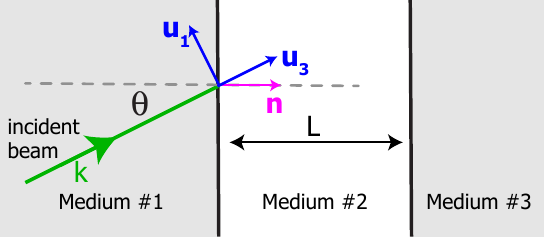}
\end{center}
\caption{{\it Double-interface geometry.}}
\label{geometry}
\end{figure}

While longitudinal and lateral centroid shifts at single interfaces have been studied for forty years, a full understanding of their nature has emerged only in the last decade~\cite{Bliokh_PRL_2006,Bliokh_PRE_2007}. As detailed below, interest in the topic has recently blossomed because of the emerging technological importance of light that carries OAM.

A light beam with OAM twists about its central axis of propagation~\cite{Nye_1974}, creating a point of zero intensity in the center that causes them to be referred to as \emph{vortex beams} or simply \emph{twisted light}~\cite{Padgett_2004, Dennis_2009}. Vortex beams can be used to manipulate spin particles encapsulated within laser traps~\cite{friese1998optical} and Bose-Einstein condensates~\cite{kapale2005vortex}, carry out micro-frabrication~\cite{omatsu2010metal}, provide control for spintronics~\cite{quinteiro2014light}, and for many other applications in which a high-fidelity torque is useful~\cite{andrews2008structured}. The quantized OAM also amounts to an additional degree of freedom for packing light with information content, so studies of OAM dynamics at interfaces are particularly relevant to new applications that exploit this in communications, computing and quantum crytography~\cite{Marrucci_2006, Yao_2011, Xi_2014, Andrews_2014,Ren_2016, Mirhosseini_2015}. 

An immediate extension of the original focus on centroid shifts at a single interface is to consider three-layer systems in which two interfaces are separated by a distance, L, as shown in Figure \ref{geometry}. For instance, Kaiser et al.~\cite{Kaiser_1996} used linearly polarized beams with no OAM to experimentally measure the GH shift that occurs when light travels through glass and is totally reflected at a thin film with a lower refractive index. It was discovered that there are angles of incidence for which the GH effect is significantly enhanced. A similar, supporting experiment was later carried out by Pillon et al.~\cite{Pillon_2005}.

GH shifts have also been theoretically considered within a different setting by Taya et al.~\cite{Taya_2012}. If the middle of three layers is of a higher refractive index than those bordering it, guided modes are supported in which GH shifts occur at each boundary.  These shifts were found to be very sensitive to the refractive indices, suggesting that such displacements could be used to measure the refractive index of the boundary material. These researchers also limited their study to linearly polarized waves without OAM. It was also noted that the shifts diminish as the width of the middle layer, L, is reduced. 

Still within the setting of linearly polarized beams without OAM, a new type of longitudinal shift was identified by Li and Wang using a combination of experimental measurements and computational modeling~\cite{Li_CPL_2004, Li_PRE_2004}. Two glass prisms were separated by a narrow air gap, and beams traveling through the left prism were incident on the first interface close to but below the critical angle. Resonant interference effects generated by the second interface resulted in what was referred to as a \emph{Generalized GH} shift that was attributable to a complex-valued reflection coefficient but not associated with evanescence. Gaussian beams encountering a multi-layer structure were also found to undergo a change in the beam waist~\cite{Tamir_1986}. 

SHEL shifts have also been explored, but to a lesser extent, in multilayer structures. Pillon et al.~\cite{Pillon_2005} predicted that there should be a spin-dependent, lateral beam shift for light traveling through glass and incident on a thin film with a lower refractive index. However, their experimental data did not match well with their model which was based on energy conservation arguments that have since been determined to contain unjustified simplifications~\cite{Yasumoto_1983, Chen_2009}. Later Menzel et al.~\cite{Menzel_2008} noted that linearly polarized beams should exhibit a small lateral shift because such beams can be decomposed into components with opposite circular polarization, and the SHEL shifts are different for each. These shifts were later computationally quantified to be on the order of tens of nanometers~\cite{Luo_PRA_2011}.

Centroid shifts have also been investigated in nontraditional layered structures. For instance, lateral shifts were computationally identified at the interface of defect layers in photonic crystals~\cite{Wang_2006}, and Dirac points in the electronic structure of graphene have been predicted to have an interesting influence on both GH and IF shifts~\cite{Chen_2009, Grosche_2015}.

Motivated by these works, we have analyzed the interaction of LG beams with two parallel interfaces separated by a distance, $L$. Our focus is on how angular momentum transfers, and in particular OAM, are affected by a second interface and manifested as lateral shifts of the reflected beam at the first interface. We also consider GH shifts to provide a unifying perspective on double-interface centroid shifts in general. Beams are assumed to have a waist that is large compared to their wavelength, so angular GH and IF shifts can be neglected. Two types of dielectric stacking are considered: low/high/low indices of refraction and high/low/high index stacking. These will be referred to as, respectively, \emph{glass sheets} and \emph{air gaps}.

We start by offering a theoretical approach that can be applied to geometries with both one and multiple interfaces. This is applied to generate salient results associated with single interfaces which allows most of the terminology and theoretical constructs to be introduced within a familiar setting. In addition, the single-interface results provide asymptotes for the multi-layer analyses that follow. A second interface is then introduced and shown to result in qualitatively new behaviors.
 
\section{Theoretical Approach}

Consider a beam characterized by 
\begin{equation}
{\bf E}({\bf r}, t) = E({\bf r}) {\bf f} e^{\imath({\bf k}_0\cdot{\bf r} - \omega t)}
\end{equation}
where ${\bf f}$ is the fixed polarization of the beam, ${\bf k}_0$ is its central wave vector, $\omega$ is the temporal frequency, $t$ is time, and the electric field strength, $E$, is a function of position, ${\bf r}$.  Since the beam is monochromatic, $|{\bf k}_0| \equiv k_0 = \omega/c$ with $c$ the speed of light in the medium through which the beam travels. A modal decomposition then exists\cite{Mandel_1995} in which the beam is described by a set of mono-length wave vectors, ${\bf k} = k_0\boldsymbol{\kappa}$. This allows direction-dependent reflection coefficients to be applied to each plane-wave mode with the results summed to obtain the reflected beam.  

It is convenient to describe the associated vector fields using beam basis vectors wherein the central axis of the beam is the unit vector, ${\bf u}_3$, as shown in Figure \ref{geometry}. Then the allowable wave vectors can be described with two independent wave numbers, $\kappa_1$ and $\kappa_2$:
\begin{equation}
\boldsymbol{\kappa} = (\kappa_1 {\bf u}_1 + \kappa_2 {\bf u}_2+\sqrt{1- \kappa_1^2-\kappa_2^2}\, {\bf u}_3)
\end{equation}
where
\begin{equation}
{\bf u}_2 = \frac{{\bf n} \times \bf{u}_3}{\bigl|{\bf n} \times \bf{u}_3\bigr|}, \quad
{\bf u}_1 = {\bf u}_2 \times \bf{u}_3.
\end{equation}
Here $\bf n$ is the unit normal to the interface as shown in Figure \ref{geometry}. Such beams have the following modal decomposition,
\begin{equation}
{\bf E}({\bf r}, t) = {\bf f} \int\!\!\!\! \int_{\mathbb{S}^2} d\kappa_1 d\kappa_2  \tilde E(\kappa_1,\kappa_2)e^{\imath(k_0\boldsymbol{\kappa}\cdot{\bf r} - \omega t)},
\end{equation}
where $\mathbb{S}^2$ is a disk of unit radius. Since only two spatial frequencies are involved, this is not a Fourier decomposition. Note that each plane wave has the same polarization as the beam. A small component of the polarization thus points along the axis of propagation of each plane wave, but this can be disregarded within the paraxial approximation. This simplifies the analysis while still capturing the salient features of all beams including the character of their angular momenta. Another form of modal decomposition is possible in which the polarization vector is decomposed into TE and TM components with respect to the propagation axis of each plane wave. As argued at length by Bliokh~\cite{Bliokh_PRE_2007}, this is not physically achievable. Moreover, working with a set of non-inertial frames does not properly capture the geometric phase that results in SHEL shifts. 

Within the paraxial approximation, a single plane-wave element of the incident beam can be expressed in the beam frame and also in the frame of the plane wave itself:
\begin{equation}
\tilde {\bf E} =   \tilde E(f_1 {\bf u}_1 + f_2 {\bf u}_2) =\tilde E_{1'} {\bf u}_{1'} + \tilde E_{2'} {\bf u}_{2'} + \tilde E_{3'} {\bf u}_{3'} .
\label{E_2ways}
\end{equation}
Here the frame of the plane wave is constructed as:
\begin{equation}
{\bf u}_{3'} = \boldsymbol{\kappa}, \quad {\bf u}_{2'} = \frac{{\bf n} \times \boldsymbol{\kappa}}{\bigl|{\bf n} \times \boldsymbol{\kappa}\bigr|}, \quad
{\bf u}_{1'} = {\bf u}_{2'} \times \boldsymbol{\kappa} .
\end{equation}
Plane waves with ${\bf u}_{1'}$ polarization are transverse magnetic (TM), and those with ${\bf u}_{2'}$ polarization are transverse electric (TE). After linearizing with respect to $\kappa_1$ and $\kappa_2$, the relationship between basis sets is
\begin{eqnarray}
{\bf u}_{1'} &=& {\bf u}_1 + \kappa_2 \mathrm{Cot}(\theta) {\bf u}_2 -\kappa_1 {\bf u}_3 \nonumber \\
{\bf u}_{2'} &=& -\kappa_2 \mathrm{Cot}(\theta) {\bf u}_1 + {\bf u}_2 -\kappa_2 {\bf u}_3 \label{primes}\\
{\bf u}_{3'} &=& \kappa_1 {\bf u}_1 + \kappa_2  {\bf u}_2 + {\bf u}_3. \nonumber
\end{eqnarray}
Here $\theta$ is the angle of incidence of the center of the beam as shown in Figure \ref{geometry}. Equations \ref{primes} can be inverted and the results used to replace ${\bf u}_1$ and $ {\bf u}_2$ in Equation \ref{E_2ways}. This allows the following expressions to be derived for $\tilde E_{1'}$ and $\tilde E_{2'}$:
\begin{eqnarray}
\tilde E_{1'} &=&\tilde E\bigl[f_1 + f_2 \kappa_2\mathrm{Cot}(\theta) \bigr]\nonumber \\
\tilde E_{2'} &=&\tilde E\bigl[f_2 - f_1 \kappa_2\mathrm{Cot}(\theta)\bigr].
\end{eqnarray}
The third component, $\tilde E_{3'} $, is of higher order in the components of $\boldsymbol{\kappa}$ and is thus disregarded, as previously anticipated. 

The $\kappa$-dependent reflection coefficients, $R^{\rm{(TM)}}$ and $R^{\rm{(TE)}}$, can now be applied to $\tilde E_{1'}$ and $\tilde E_{2'}$, respectively. Then the primed basis vectors can be once again re-expressed in terms of the basis vectors of the beam using Equation \ref{primes}. This gives an expression for each plane wave element of the reflected beam:
\begin{eqnarray}
\tilde {\bf E}_{R} =&&\tilde E\biggl\{\bigl[f_1 R^{\rm{(TM)}} - f_2\kappa_2 \mathrm{Cot}(\theta)(R^{\rm{(TM)}}+R^{\rm{(TE)}})\bigr]{\bf u}_1 \nonumber \\
+&&\bigl(f_2 R^{\rm{(TE)}} + f_1\kappa_2 \mathrm{Cot}(\theta)(R^{\rm{(TM)}}+R^{\rm{(TE)}})\bigr){\bf u}_2 \\
-&& \bigl[f_1 R^{\rm{(TM)}}\kappa_1 +f_2 R^{\rm{(TE)}}\kappa_2 \bigr]{\bf u}_3 \biggr\} .\nonumber
\end{eqnarray}
Since the beam basis vectors are constant, the reflected beam is easily recovered by summing over all of its components:
\begin{equation}
{\bf E}_{R} ({\bf r},t) = \int\!\!\!\! \int_{\mathbb{S}^2} d\kappa_1 d\kappa_2  \tilde {\bf E}_{R}(\kappa_1,\kappa_2) e^{\imath(k_0\boldsymbol{\kappa}\cdot{\bf r} - \omega t)}.
\label{ER}
\end{equation}

For a single interface, the standard Fresnel coefficients are used for $R^{\rm{(TM)}}$ and $R^{\rm{(TE)}}$:
\begin{equation}
R^{\rm{(TM)}}= \frac{\mathrm{Cos}(\beta) - C}{\mathrm{Cos}(\beta) - C},\quad
R^{\rm{(TE)}} = \frac{n^2 \mathrm{Cos}(\beta) - C}{n^2 \mathrm{Cos}(\beta) - C}.
\label{fresnelsingle}
\end{equation}

For double-interface settings, the appropriate reflection coefficients are easily derived to be
\begin{widetext}
\begin{eqnarray}
R^{\rm{(TM)}}&=& -\frac{(-1+\Phi ) \left(-n^2+n^4 \text{Cos}(\beta )^2+\text{Sin}(\beta )^2\right)}{-2 C n^2 (1+\Phi ) \text{Cos}(\beta )+n^4 (-1+\Phi ) \text{Cos}(\beta
)^2+(-1+\Phi ) C^2} \nonumber \\
R^{\rm{(TE)}} &=& -\frac{\left(-1+e^{i 2 L C^2}\right) \left(-1+n^2\right)}{-2 C (1+\Phi ) \text{Cos}(\beta )+(-1+\Phi )
\text{Cos}(\beta )^2+(-1+\Phi ) C^2}.
\label{fresneldouble}
\end{eqnarray}
\end{widetext}
In both cases, the following short-hand has been introduced:
\begin{eqnarray}
\beta:=\mathrm{Cos}^{-1}(\boldsymbol{\kappa}\cdot {\bf n}) \nonumber \\
C := \sqrt{n^2 - \mathrm{Sin}^2(\beta)}\\
\Phi := e^{\imath 2 L C} . \nonumber
\label{beta}
\end{eqnarray}

Longitudinal (GH) and lateral (SHEL/OIF) shifts in the reflected beam centroids are defined, respectively, as
\begin{eqnarray}
\langle u_1 \rangle &=&  \frac{\int\!\!\!\! \int_{\mathbb{R}^2} du_1 du_2  |{\bf E}_{R} ({\bf r})|^2 u_1}{ \int\!\!\!\! \int_{\mathbb{R}^2} du_1 du_2  |{\bf E}_{R} ({\bf r})|^2} \nonumber \\
\quad \label{centroids}\\
\langle u_2 \rangle &=&  \frac{\int\!\!\!\! \int_{\mathbb{R}^2} du_1 du_2  |{\bf E}_{R} ({\bf r})|^2 u_2}{ \int\!\!\!\! \int_{\mathbb{R}^2} du_1 du_2  |{\bf E}_{R} ({\bf r})|^2}. \nonumber
\end{eqnarray}
where position vector ${\bf r} = \sum_{n=1}^3 u_n {\bf u_n}$. 

For general LG beams, the $\kappa_1$, $\kappa_2$ plane is discretized and Fast Fourier transforms are used to carry out the modal decomposition and reconstruction, while numerical integration is used to quantify the centroid shifts above.

\subsection{Analytical Expressions for Gaussian Beams}
For the special case of Gaussian beams, it is possible to analytically evaluate the shifts of Equation \ref{centroids} within the frequency domain using the following equivalent centroid definitions:
\begin{eqnarray}
\langle u_1 \rangle&=&  \frac{\imath \int\!\!\!\! \int_{\mathbb{R}^2}d\kappa_1 d\kappa_2 \frac{\partial {\tilde{\bf E}}_{R}}{\partial \kappa_1}\cdot \tilde {\bf E}_{R}^*}{\int\!\!\!\! \int_{\mathbb{R}^2}d\kappa_1 d\kappa_2  |\tilde{\bf E}_{R} |^2}\nonumber \\
\quad \\
\langle u_2 \rangle&=&  \frac{-\imath \int\!\!\!\! \int_{\mathbb{R}^2}d\kappa_1 d\kappa_2 \frac{\partial {\tilde{\bf E}}_{R}}{\partial \kappa_2}\cdot \tilde {\bf E}_{R}^*}{\int\!\!\!\! \int_{\mathbb{R}^2}d\kappa_1 d\kappa_2  |\tilde{\bf E}_{R} |^2}. \nonumber
\label{centroidfreq}
\end{eqnarray}

To carry this out, we approximate, to first order in $\kappa_1$ and $\kappa_2$, the angle of incidence, $\beta$, of individual plane wave elements given in Equation \ref{beta}:
\begin{equation}
\beta = \theta + \kappa_1 + O(\kappa_1^2) + O(\kappa_2^2) .
\label{betalin}
\end{equation}
Within this linearized setting, the reflection coefficients do not depend on $\kappa_2$ and are linear functions of $\kappa_1$:
\begin{equation}
R^{\rm{(TM)}} = R^{\rm{(TM)}}_{0} + \kappa_1 R^{\rm{(TM)}}_{1}, \quad R^{\rm{(TE)}} = R^{\rm{(TE)}}_{0} + \kappa_1 R^{\rm{(TE)}}_{1} .
\label{Rlin}
\end{equation}
This is true for both single- and double-interfaces allowing expressions to be derived that are common to both settings. In particular, assuming that the beam waist is large compared to its wavelength, it is not hard to show that the GH shifts are well-approximated by:
\begin{eqnarray}
\langle u_1 \rangle_{\rm TM} &=&  \frac{1}{k_0 }{\rm Im}\biggl(\frac{R^{\rm{(TM)}}_{1}}{R^{\rm{(TM)}}_{0}}\biggr),\quad \langle u_1 \rangle_{\rm TE} = \frac{1}{k_0 }{\rm Im}\biggl(\frac{R^{\rm{(TE)}}_{1}}{R^{\rm{(TE)}}_{0}}\biggr) \nonumber \\
\langle u_1 \rangle_{\pm} &=&  \frac{{\rm Im}\bigl(R^{\rm{(TM)}}_{1} R^*_{10}+ R^{\rm{(TE)}}_{1}R^*_{20}\bigr)}{k_0\bigl(\big|R^{\rm{(TM)}}_{0}\big|^2 + \big| R^{\rm{(TE)}}_{0}\big|^2\bigr)} .
\label{centroidx}
\end{eqnarray}

Likewise, the SHEL shifts are given by:
\begin{equation}
\langle u_2 \rangle_\pm =  \frac{\mp \big|R^{\rm{(TM)}}_{0}+ R^{\rm{(TE)}}_{0}\big|^2 \mathrm{Cot}(\theta)}{k_0\bigl(\big|R^{\rm{(TM)}}_{0}\big|^2 + \big|R^{\rm{(TE)}}_{0}\big|^2\bigr)} .
\label{centroidy}
\end{equation}

These expressions are particularly useful because they can be immediately applied to single interfaces using Equation \ref{fresnelsingle} and double-interfaces using Equation \ref{fresneldouble}. The single-interface application recovers Artmann's formulae~\cite{Artmann_1948} for GH shifts and the SHEL shift predictions of Bliokh\cite{Bliokh_PRL_2006,Bliokh_PRE_2007}, and these are equivalent to more general expressions derived elsewhere~\cite{Bliokh_2013}. 
%
%

\section{Brief Summary of Single-Interface Shifts}

Within a paraxial beam approximation, we restrict attention to LG beams that travel through air(glass) and are incident on glass(air). The ratio of refractive indices, right medium divided by left medium, will be denoted by $n$. There are three types of centroid shifts possible: (1) GH shifts for $n<1$ and a super-critical angle of incidence; (2) SHEL shifts due to SAM for all values of $n$ and angles of incidence; and (3) OIF shifts due to OAM for all values of $n$ and angles of incidence. We quantify each of these using borosilicate glass Schott BK7, $(n = 1.5168)$. The product of wavenumber, $k_0$, to beam waist, $w_0$ was set to a value of $2\times 10^4$.

\subsubsection{Air/Glass Interface}

\begin{figure}[hptb]
\begin{center}
\includegraphics[width=0.4\textwidth]{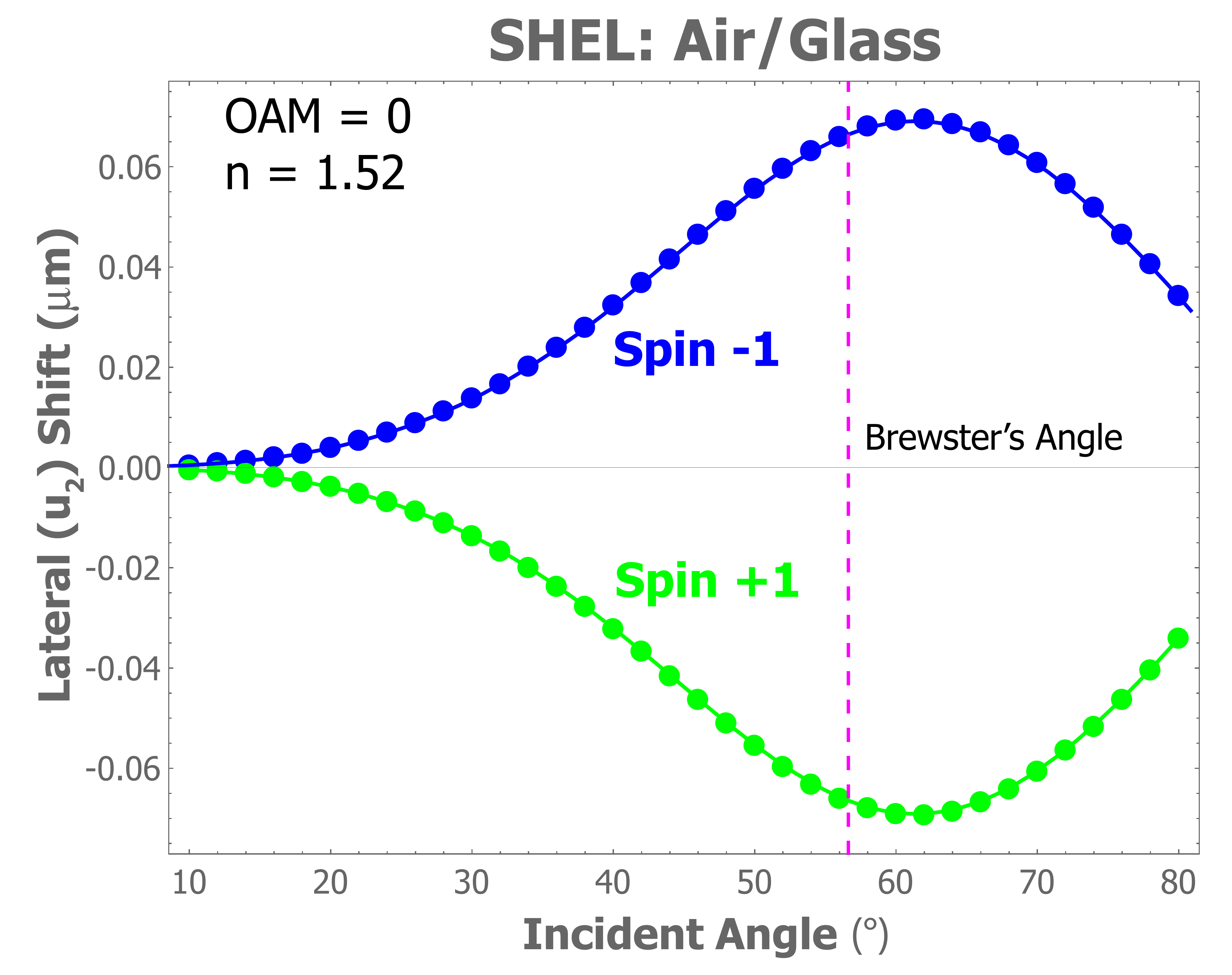}
\end{center}
\caption{{\it Centroid shifts for a single air/glass interface.}  OAM = 0. SHEL shifts generated by circularly-polarized beams. Discrete points are computational data while the smooth curves are the sub-critical, analytical expressions of  Bliokh~\cite{Bliokh_PRL_2006}. Brewster's angle, $56.8^\circ$, is shown by a magenta dotted line. }
\label{SingleInterface_airglass_NO_OAM}
\end{figure}

When light is incident on a medium of higher dielectric constant, there is no critical angle for total internal reflection--i.e. the reflection coefficient is real-valued--implying that there is no GH shift. The SHEL response associated with circularly-polarized light is shown in Figure \ref{SingleInterface_airglass_NO_OAM}. There and henceforth, right(left) circularly-polarized beams will be referred to as \emph{spin + (spin -)} since they correspond to spin quantum numbers in quantum theory. Note that these shifts very small ($<10 nm$) and are not sensitive to the Brewster angle. Since linearly polarized beams can be decomposed into spin components, they also have a lateral shift, but it is orders of magnitude smaller because it relies on the asymmetry in the shifts for each type of spin~\cite{Hosten_2008}.

%
%

\begin{figure}[hptb]
\begin{center}
\includegraphics[width=0.48\textwidth]{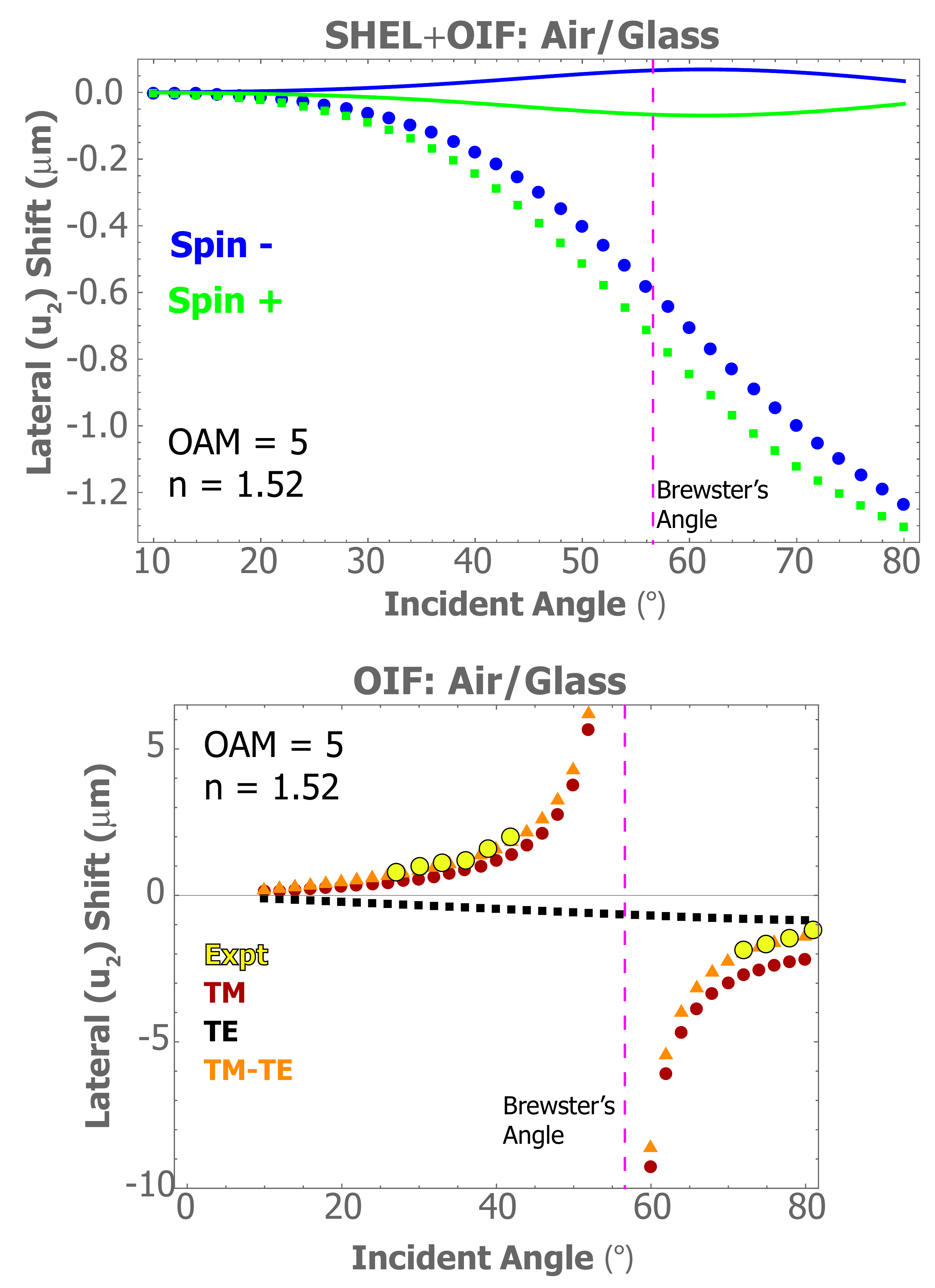}
\end{center}
\caption{{\it SHEL and OIF shifts for a single air/glass interface.}  OAM = 5. Top plot shows that both SHEL and OIF contribute to the lateral shift of circularly-polarized vortex beams. The solid curves in the top plot are from the analytical expression of  Bliokh~\cite{Bliokh_PRL_2006} for the SHEL shifts, and the difference between these curves and the discrete points represents the OIF contribution. Bottom plot shows OIF shift of TM-polarized beams and the singular response at Brewster's angle of $56.8^\circ.$ Experimental data is from Dasgupta~\cite{Dasgupta_2006}.}
\label{SingleInterface_airglass}
\end{figure}

The addition of OAM to the beam generates an OIF response that can be observed in TM-polarized beams as shown in the bottom plot of Figure \ref{SingleInterface_airglass}. This shift becomes singular at the Brewster angle, $56.8^\circ$, where only the vanishing strength of the reflected beam effectively limits the size of the OIF response that can be measured.  The lateral shifts of circularly-polarized beams are now due to a combination of SHEL and OIF as shown at top in Figure \ref{SingleInterface_airglass} and should be compared directly to the plot of Figure \ref{SingleInterface_airglass_NO_OAM} to see the effect of OAM on the shifts. 

%
%

\subsubsection{Glass/Air Interface}
\begin{figure}[hptb]
\begin{center}
\includegraphics[width=0.42 \textwidth]{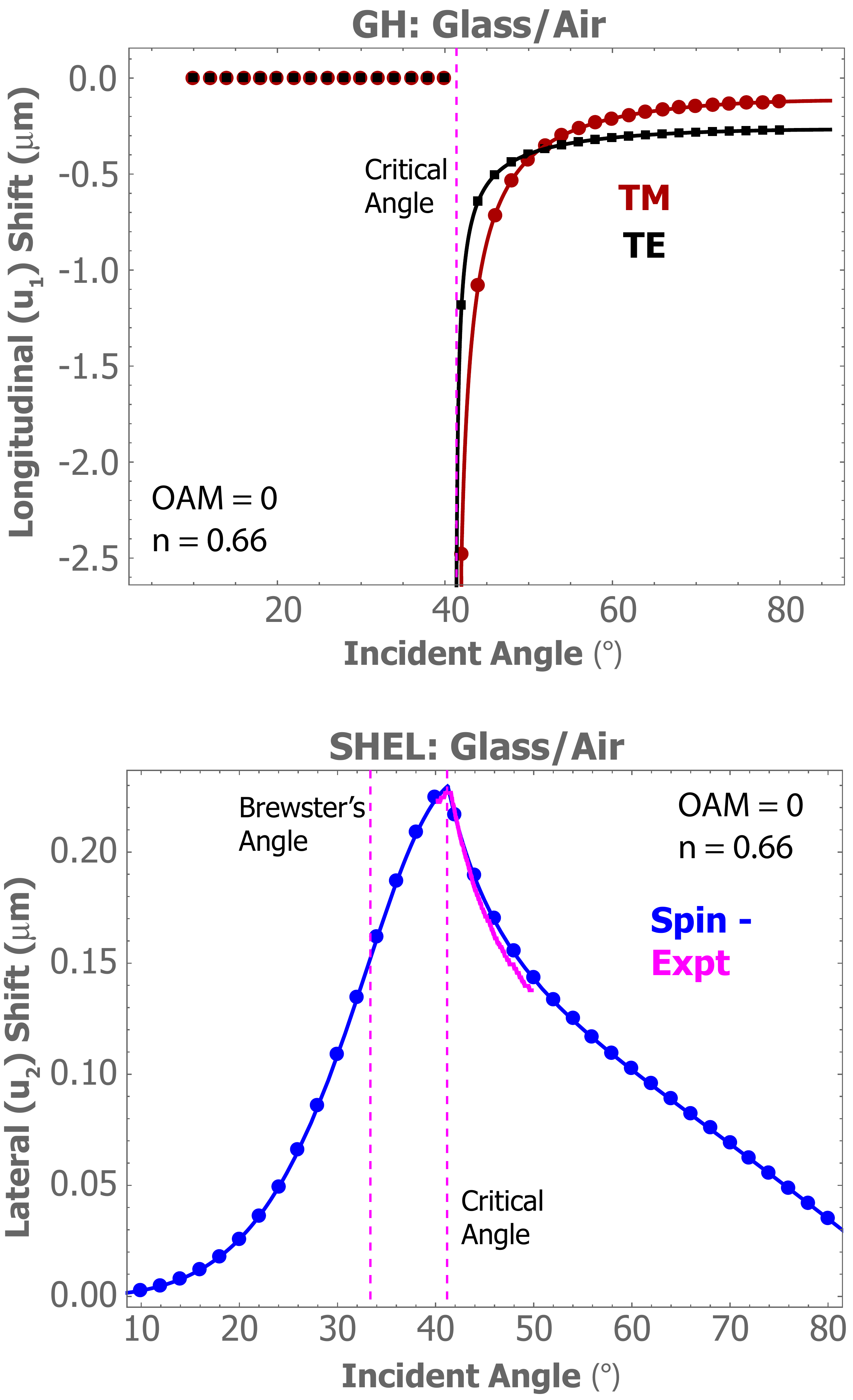}
\end{center}
\caption{{\it GH and SHEL shifts for a single glass/air interface.}  OAM = 0. The discrete points are computational data while the smooth curves are analytic approximations: Artmann's formula~\cite{Artmann_1948} for the GH shift and Bliokh's formulae~\cite{Bliokh_PRL_2006} for the SHEL shifts. Experimental data (bottom panel, magenta) is from Dasgupta~\cite{Dasgupta_2006}. Brewster's angle, $33.4^\circ$, and the critical angle, $41.2^\circ$, are shown by magenta dotted lines. Note that the top plot is for linear polarizations while the bottom plot is for circularly-polarized beams.}
\label{SingleInterface_glassair_NO_OAM}
\end{figure}

Now consider a LG beam traveling through glass that is incident on a planar interface with air. Below the critical angle, $41.2^\circ$, there is no GH shift since the reflection coefficient is real-valued. At the critical angle, the GH shift is singular, decreasing rapidly for larger angles of incidence as shown at top in Figure \ref{SingleInterface_glassair_NO_OAM}. In the absence of OAM, circularly-polarized light exhibits a SHEL shift at all angles of incidence as shown in the bottom plot of Figure \ref{SingleInterface_glassair_NO_OAM}.  As with the Air/Glass interface, the lateral shifts of linearly polarized beams are orders of magnitude smaller because they rely on the asymmetry in the shifts for each type of spin. 

%
%

\begin{figure}[hptb]
\begin{center}
\includegraphics[width=0.42\textwidth]{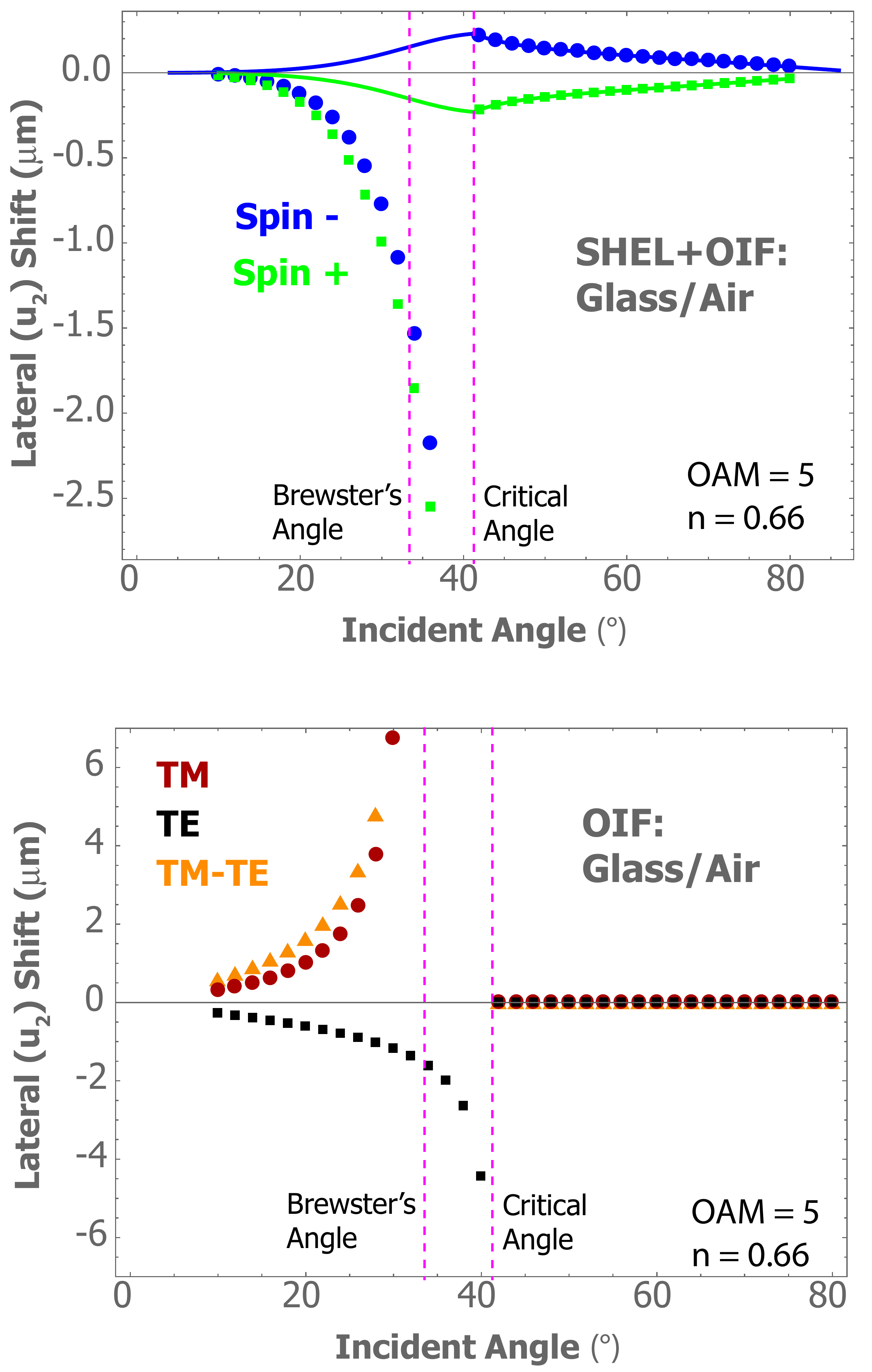}
\end{center}
\caption{{\it Centroid shifts for a single glass/air interface.}  OAM = 5. The critical angle is at $41.2^\circ.$ Top plot shows combined SHEL/OIF shifts for circularly-polarized vortex beams, where the singular response at the critical angle is due to the OIF contribution. The solid curves are from Bliokh's formulae~\cite{Bliokh_PRL_2006} for the SHEL shift, and the departure from these curves represents the OIF contribution. Bottom plot shows the OIF shift associated with linearly polarized beams. Brewster's angle is $56.8^\circ.$}
\label{SingleInterface_glassair}
\end{figure}

The inclusion of OAM does not affect the GH shifts, but the interface causes such beams to distort from simple vortices. This converts some of the intrinsic OAM to an extrinsic OAM. TM-polarized vortex beams will  exhibit a strong lateral shift near the Brewster angle, as shown at bottom in Figure \ref{SingleInterface_glassair}.  This is always in the sub-critical range because the Brewster angle is less than the critical angle. Only the vanishing strength of the reflected beam puts an effective limit on the size of the lateral response that can be measured. The lateral shifts of circularly-polarized beams are now due to a combination of SHEL and OIF, as shown in the top plot in Figure \ref{SingleInterface_glassair} which can be compared directly to the bottom plot of Figure \ref{SingleInterface_glassair_NO_OAM} to see the effect of OAM on the shifts. 

%
%

In summary, our numerical results for single-interface shifts are well validated by the combination of our own analytical results and the experimental and analytical results of others. Since the same computational machinery is used in the consideration of double-interfaces, this lends confidence to the new results obtained in that setting.

\section{Results for Double-Interfaces}

We now turn to a consideration of how GH and IF shifts are influenced by the addition of a second interface. As before,  a Laguerre-Gaussian (LG) beam travels through the left vacuum and is incident on the left interface. Two geometries are considered. The first is that of a thin sheet of glass with air on either side in a \emph{glass sheet} geometry. The second is just the opposite, two semi-infinite glass prisms sandwiching a thin gap, an \emph{air gap} geometry. 

As discussed above, TM-polarized plane waves incident on a single interface at the Brewster angle have no reflected component, and this implies that TM-polarized LG beams will have a very low reflected signal when their central axis is at the Brewster angle. We have noted that the IF shift is effectively singular at this angle. Since this only holds for TM-polarized light, the difference in reflected centroid position between TE and TM beams incident near the Brewster angle provides a nice way of quantifying IF shifts. 

It turns out that there is an analogous, tunable behavior associated with double interfaces. For a prescribed angle of incidence and refractive index, there exist a countably infinite set of sheet thicknesses for which the reflection coefficient of plane waves is zero due to destructive interference of beams reflected from both interfaces. In the absence of OAM, it is clear from the expressions of Equations \ref{centroidx} and \ref{centroidy} that, for finite-valued numerators, centroid shifts will be singular at such points. The addition of a beam vortex, though, must be treated numerically. In general, a condition for which the reflection coefficient becomes zero because of multi-layer interference will be referred to as a \emph{Fabry-Perot Resonance} (FBR) in the centroid shifts. These will be elucidated below with an eye towards unifying previous reports and extending the consideration to vortex beams. 

\subsection{Beams with No OAM}

\begin{figure}[hptb]
\begin{center}
\includegraphics[width=0.45\textwidth]{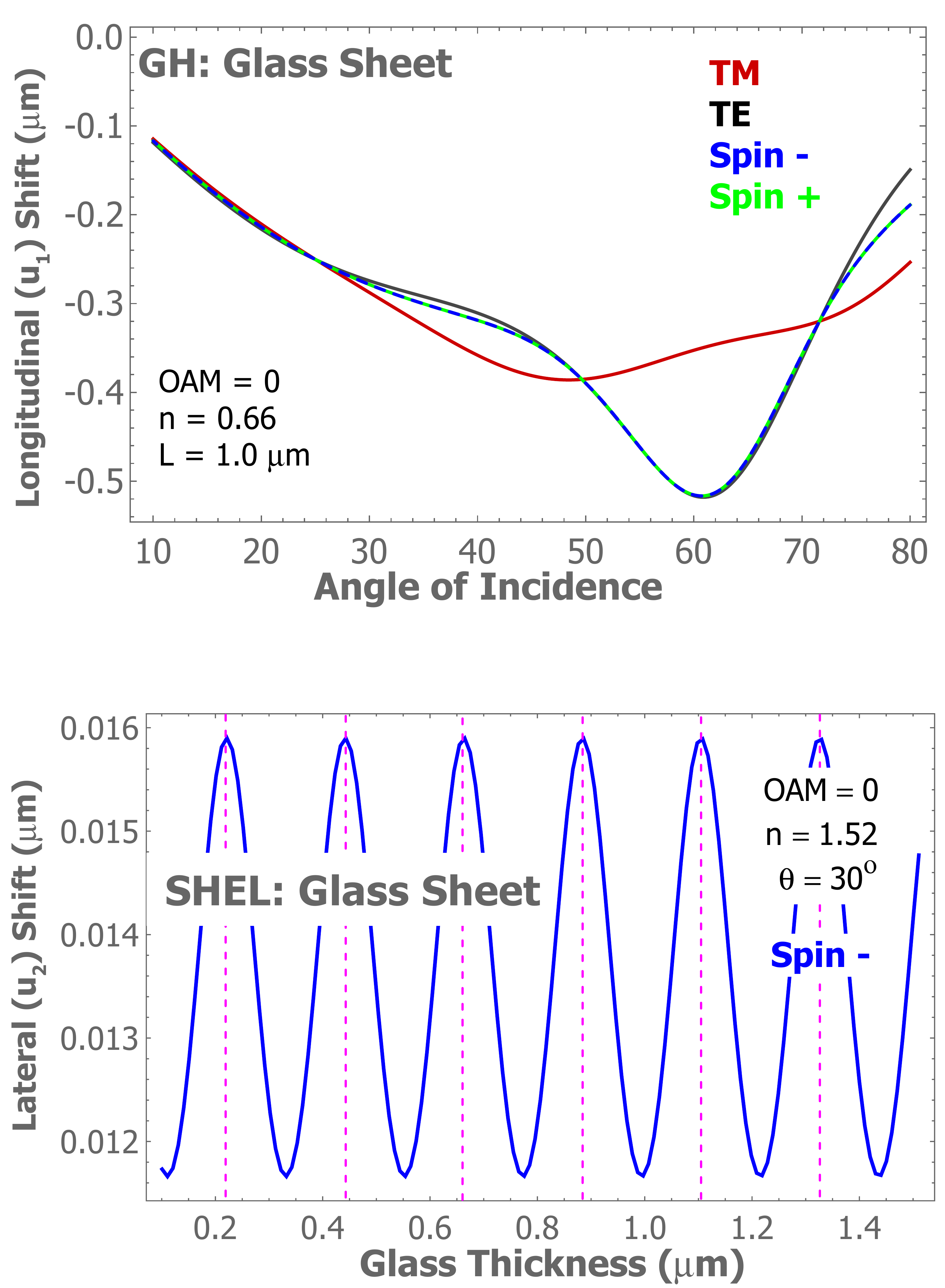}
\end{center}
\caption{{\it Centroid shifts for a glass sheet.} Top plot shows Generalized GH shifts~\cite{Li_PRE_2004}.  L = $1.0  \mu m$, $OAM = 0$. The curves are generated from the analytical formulae of Equations \ref{fresneldouble} and \ref{centroidx}. Bottom plot shows the thickness dependence of the SHEL shift of a circularly-polarized beam for an angle of incidence = $30^\circ$, $OAM = 0$. The dashed lines are the predicted locations of FBR. The curves are generated from the analytical formulae of Equations \ref{fresneldouble} and \ref{centroidy}. Beams with the opposite circular polarization exhibit a shift of the opposite sign. }
\label{Double_Interface_Glass_Sheet_Summary_No_OAM}
\end{figure}

Figure \ref{Double_Interface_Glass_Sheet_Summary_No_OAM} shows results for beams with no OAM incident on a glass sheet. The second interface causes the reflection coefficients to be complex-valued, resulting in Generalized GH shifts--i.e. longitudinal shifts not associated with an evanescent component in the middle layer~\cite{Li_PRE_2004}. The small SHEL shifts follow the anticipated trend in which magnitudes are greater at FBR angles. In the absence of OAM, the IF shift is due only to SHEL and is only weakly sensitive to FBR. An analysis of Equation \ref{centroidy} explains why the lateral shifts are not singular at FBR points; it is because both numerator and denominator go to zero with the ratio remaining finite. Thus, the FBR condition is necessary but not sufficient to produce singularities in centroid shifts. As was previously noted in association with single interfaces, the SHEL shift associated with linearly polarized beams is orders of magnitude smaller because it relies on the asymmetry in the shifts for each type of spin~\cite{Hosten_2008}. The same is true for double-interfaces as well~\cite{Menzel_2008}. 

%
%

\begin{figure}[hptb]
\begin{center}
\includegraphics[width=0.42\textwidth]{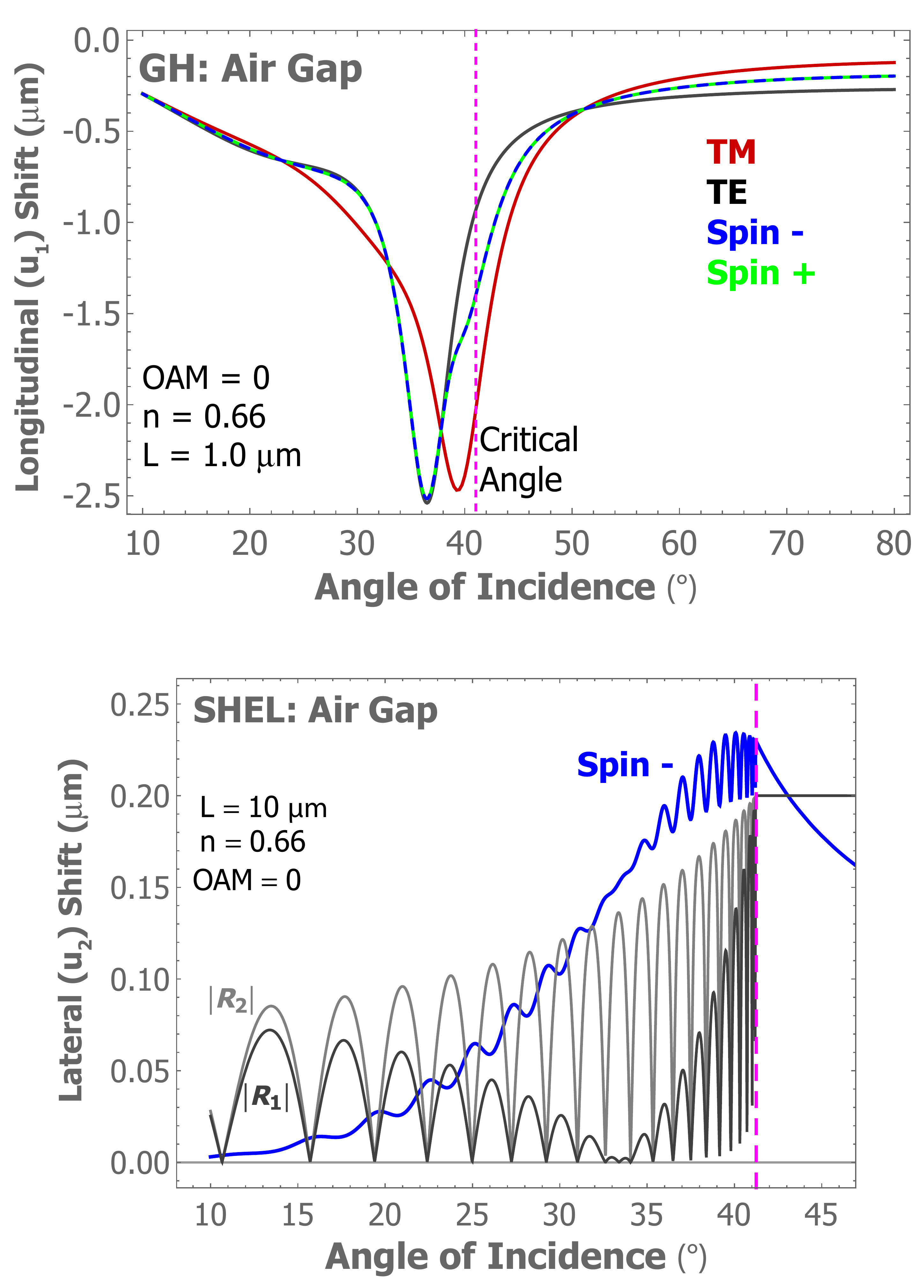}
\end{center}
\caption{{\it Centroid shifts for air gaps.} Top plot shows GH shifts. L = $1.0  \mu m$, $OAM = 0$. The curves are generated from the analytical formulae of Equations \ref{fresneldouble} and \ref{centroidy}. Bottom plot quantifies SHEL shift bunching. Width = 10.0 $\mu m$, OAM = 0. FBR results in a SHEL shift that varies rapidly as the angle of incidence is changed. The curve is generated from the analytical formulae of Equations \ref{fresneldouble} and \ref{centroidy}. Beams with the opposite circular polarization exhibit a shift of the opposite sign. There are no TE or TM shifts.}
\label{Double_Interface_Air_Gap_Summary_No_OAM}
\end{figure}

Air gaps also exhibit GH shifts over the entire range of incidence angles, where the subcritical shifts are due to the Generalized GH effect as noted in earlier work~\cite{Li_PRE_2004}. This is shown in the top plot of Figure \ref{Double_Interface_Air_Gap_Summary_No_OAM}. There is also a substantial influence of FBR on IF shifts though. As shown in the bottom plot of the same figure, beams  incident at below the critical angle ($41.2^\circ$) exhibit an oscillatory response with changing angle of incidence that becomes increasingly compressed as the critical angle is approached. Angle-dependent troughs in the lateral shifts are seen to correspond to nodes in the reflection coefficients. Such FBR bunching increases monotonically with the width of the air gap. The plot can be directly compared with that of the bottom panel of Figure \ref{SingleInterface_glassair_NO_OAM} to see the effect of the second interface. This new behavior should be readily observable near the critical angle.

%
%

\begin{figure}[hptb]
\begin{center}
\includegraphics[width=0.4\textwidth]{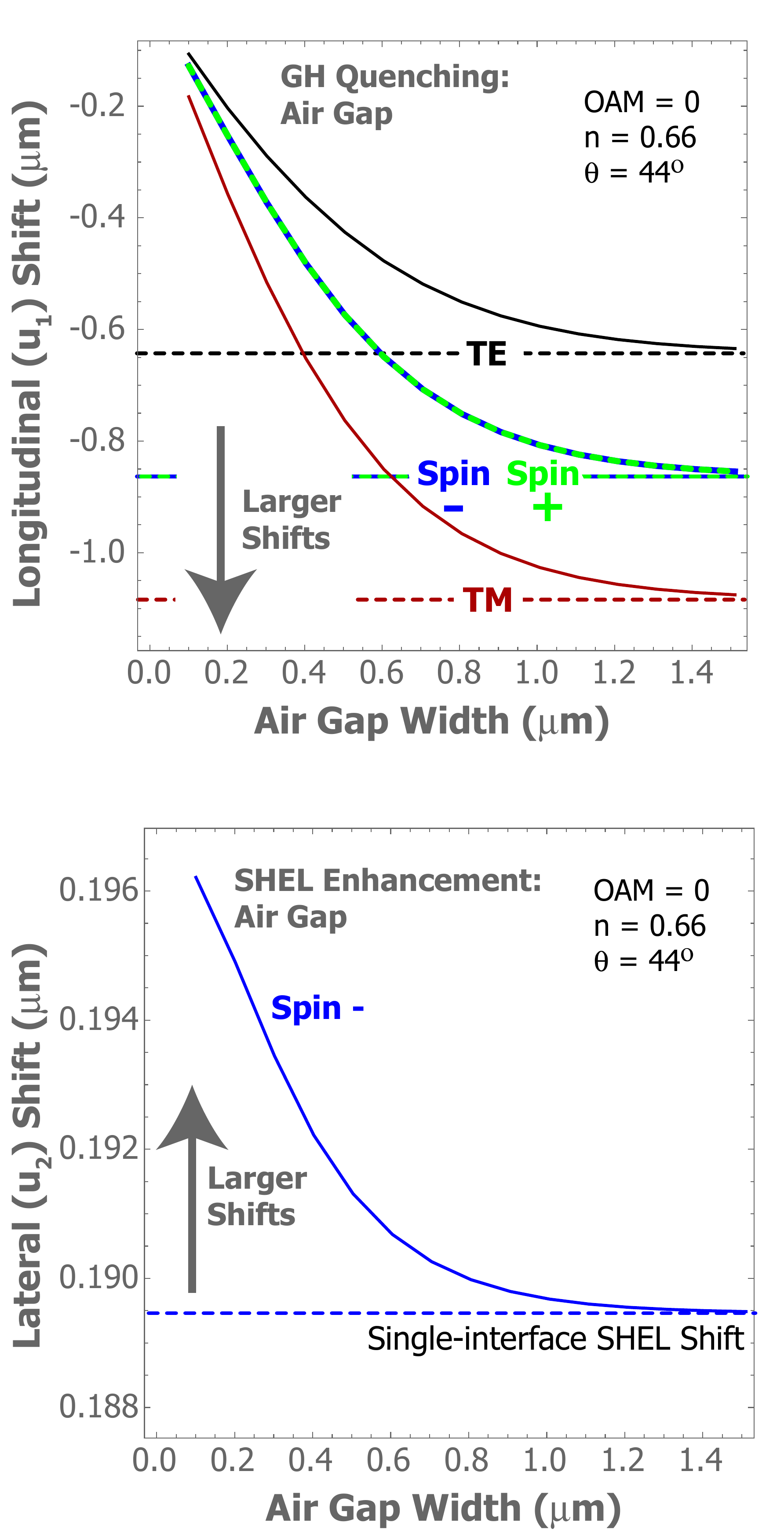}
\end{center}
\caption{{\it GH quenching and SHEL enhancement versus air gap width.} Angle of incidence = $44^\circ$, above the critical angle of $41.2^\circ$. OAM = 0. The dashed lines are the shifts associated with a single interface. The curve is generated from the analytical formulae of Equations \ref{fresneldouble} and \ref{centroidy}. The SHEL shift associated with positive circular polarization is identical in magnitude but of the opposite sign.}
\label{Double_Interface_Width_Air_Gap_No_OAM}
\end{figure}

The air gap also has a significant influence on the magnitude of the longitudinal (GH) shift for angles of incidence beyond the critical angle~\cite{Taya_2012}. In particular, the GH shift decreases with decreasing gap thickness in what might be referred to as \emph{Goos-H{\"a}nchen Quenching}. This is shown in the top plot of Figure \ref{Double_Interface_Width_Air_Gap_No_OAM}. For sufficiently thick sheets, evanescent decay will effectively decouple the dynamics of the two interfaces and the shifts associated with a single interface are recovered. This quenching behavior is the result of Frustrated Total Internal Reflection, in which a component of the beam propagates into the second prism of glass instead of tunneling along the interface. The result is a GH shift that decreases as the transmitted fraction of light increases.

The bottom panel of Figure \ref{Double_Interface_Width_Air_Gap_No_OAM} shows just the opposite behavior, albeit weak, for lateral shifts of the beam. The plot shows that SHEL shifts actually increase as the thickness of the air gap is decreased--i.e. there is a \emph{SHEL Enhancement} for both circularly-polarized beams.

%
%

\subsection{Vortex Beams}

\begin{figure}[hptb]
\begin{center}
\includegraphics[width=0.40\textwidth]{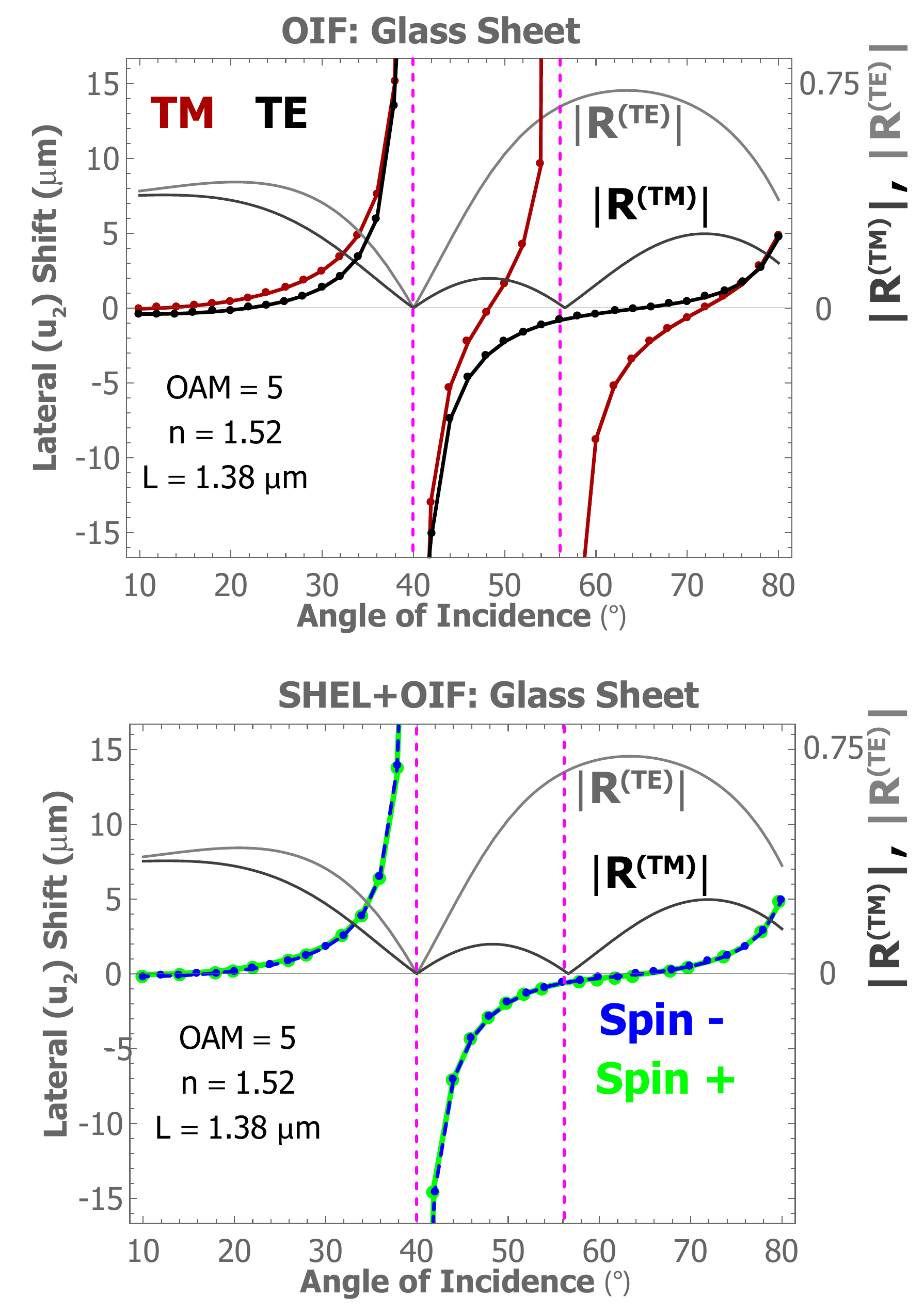}
\end{center}
\caption{{\it SHEL and OIF shifts for glass sheet.} Width = 1.38 $\mu m$, OAM = 5. There is a singular TM response at the Brewster angle, $56.8^\circ$, but the FBR generates a new singular response at $40^\circ$ as well. This second singularity appears for all 4 polarizations. Computational data is shown as discrete points with solid curves serving as a guide to the eye.}
\label{Double_Interface_Angle_Glass_Sheet}
\end{figure}

The effect of FBR on the SHEL shifts at glass sheets was found to be very mild (bottom panel of Figure \ref{Double_Interface_Glass_Sheet_Summary_No_OAM}), but this changes dramatically when OAM is included in the beam. This is demonstrated in Figure \ref{Double_Interface_Angle_Glass_Sheet}. For the sheet thickness chosen, a single FBR exists at $40^\circ$, and all four polarizations exhibit a shift singularity at that angle of incidence. On the other hand, only the TM polarized beam has a singular response at the Brewster angle, $56.8^\circ$. As is clear from the figure, there are ranges of beam incidence for which both the centroid shift and the reflection coefficients are predicted to be relatively large ($>10 \mu m$). This implies that these OIF shifts should be straightforward to measure.

%
%

\begin{figure}[hptb]
\begin{center}
\includegraphics[width=0.45\textwidth]{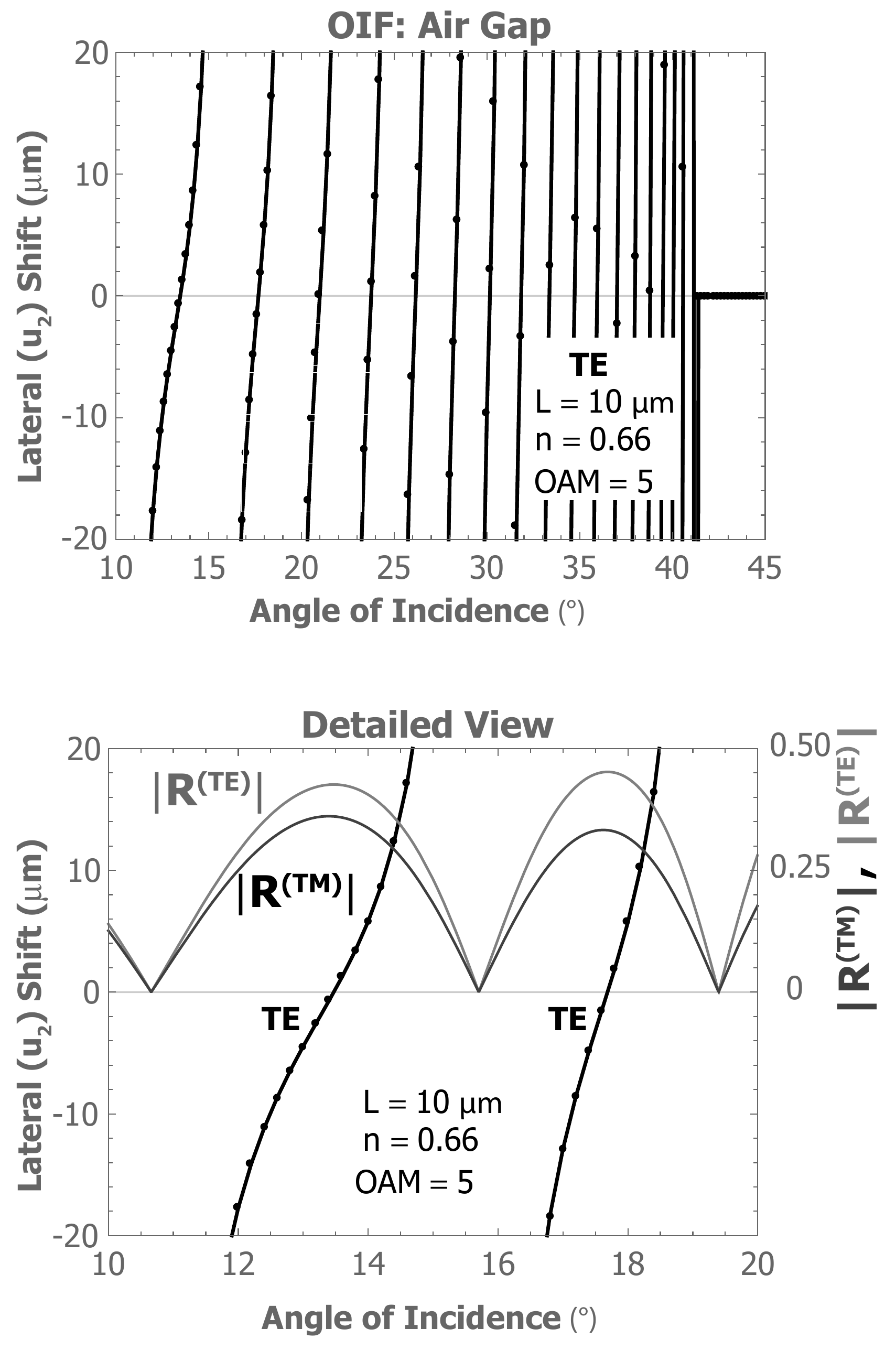}
\end{center}
\caption{{\it SHEL and OIF shifts versus angle of incidence.} Air gap thickness = $10 \mu m$, critical angle = $41.2^\circ$, OAM = 5. Top plot shows rapid oscillation in OIF shifts with angle of incidence for a TE polarized beam. Bottom plot is a zoomed view that includes the reflection coefficients. Computational data is shown as discrete points with solid curves serving as a guide to the eye. The GH shifts are not affected by OAM and so follow the same trends as shown in Figure \ref{Double_Interface_Air_Gap_Summary_No_OAM}.}
\label{Double_Interface_Width_Air_Gap}
\end{figure}

The dielectric values are now switched in order to consider an air gap between two glass prisms. FBR can be used to engineer a large OIF shift as shown in Figure \ref{Double_Interface_Width_Air_Gap}. This is visually similar to the subcritical behavior observed without OAM as shown in the lower panel of Figure \ref{Double_Interface_Air_Gap_Summary_No_OAM}, but linear polarizations are now seen to elicit the response previously associated only with circular polarizations. Data is shown only for TE-polarized beams, but the same behavior is observed for the other three polarizations as well. There are once again ranges of beam incidence for which both the centroid shift and the reflection coefficients are predicted to be relatively large ($>10 \mu m$).

%
%

\begin{figure}[hptb]
\begin{center}
\includegraphics[width=0.4\textwidth]{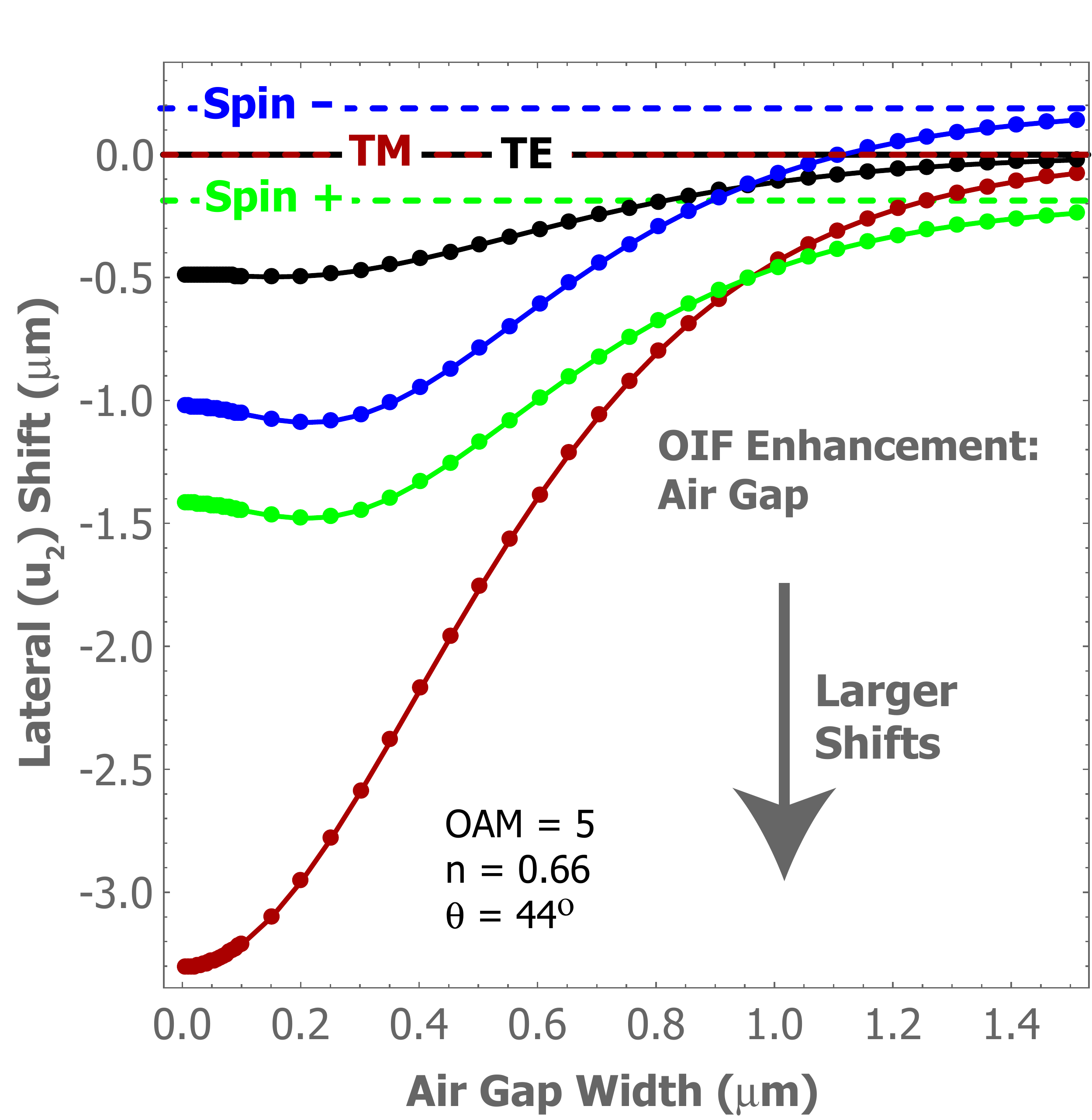}
\end{center}
\caption{{\it Enhancement of SHEL and OIF shifts versus air gap width.} Angle of incidence = $44^\circ$, critical angle of $41.2^\circ$, OAM = 5. Large lateral shifts for vortex beams as compared with the mild enhancement attributable to SAM as shown in Figure \ref{Double_Interface_Width_Air_Gap_No_OAM}. Computational data is shown as discrete points with solid curves serving as a guide to the eye. A denser set of data points were obtained for narrow widths to more clearly show trends there. Dashed lines show the single-interface shifts.}
\label{Double_Interface_Width_Air_Gap_IF_Enhancement}
\end{figure}

Finally, the inclusion of OAM fundamentally changes the enhancement of lateral beam shifts with decreasing air gap width. Figure \ref{Double_Interface_Width_Air_Gap_IF_Enhancement} quantifies this for all for polarizations, and the results should be compared directly with those for a beam with only SAM, the bottom plot of Figure \ref{Double_Interface_Width_Air_Gap_No_OAM}. It is clear that the addition of intrinsic OAM has a dominant influence on the shifts, even changing the sign of the shift from that predicted for spin polarizations of the opposite sign. This data is associated with an angle of incidence of $44^\circ$, just above the critical angle of $41.2^\circ$ where the lateral shifts are still relatively large. Because this incident angle is close to the Brewster angle, $40.8^\circ$, the TM beams show a shift enhancement over six times larger than TE beams. 

%
%

\section{Conclusions}

The interaction of light with double-interface structures has been considered in a comprehensive manner in order to unify previous results and to extend them to show how they are qualitatively changed for vortex beams. For the sake of clarity, this work has focused on beams that are wide relative to the wavelength of light, allowing angular shifts to be neglected. The response of light at material interfaces is simplified by decomposing beams into a linear combination of plane waves of the same polarization but with assumed small departures from the central wave vector direction. Shifts are quantified for reflected beams, but the effects identified will certainly have manifestations for transmitted beams as well. 

The behaviors observed can be accounted for in terms of two basic effects: Fabry-Perot Resonances for propagating modes; and coupling between interfaces for evanescent modes. The former results in generalized GH shifts and a mild modulation of SHEL shifts for glass sheets as well as subcritical GH shifts and SHEL shift bunching for air gaps. These are independent of polarization and will be observable even for linearly polarized beams. The evanescent coupling between interfaces tends to quench the GH shift as the interlayer width is decreased. Surprisingly though, the opposite trend is observed for SHEL shifts which are slightly enhanced. 

Vortex beams show behavior that is qualitatively different than their purely spin counterparts. The previously identified small modulations in SHEL shifts are now dominated by strong singularities for both glass sheets and air gaps. These should be particularly easy to observe, are independent of polarization, and should exhibit an increasingly rapid oscillation with incidence angle as the thickness of the interlayer is increased.  In addition, for beams incident on air gaps beyond the critical angle, the enhancement of lateral shifts for thin gaps is many times larger than that due to spin alone. This effect should also be straightforward to observe because the shifts are on the order of microns.

\section{Acknowedgements}
M. T. L. acknowledges helpful discussions with D. Andrews of East Anglia University. M. E. S. acknowledges support from the National Science Foundation through grants CMP-1553905 and EPMD-1509733. G. F. Q. acknowledges support from the University of Denver Marsico Scholar program. 
 
\bibliographystyle{prsty}

\end{document}